\documentclass[10pt]{iopart}

\usepackage{iopams}
\usepackage{graphicx} % Include figure files
\usepackage{dcolumn}% Align table columns on decimal point
\usepackage{upgreek }
\usepackage{times}
\usepackage{textcomp}
\usepackage{ wasysym }
\usepackage{titlesec}
\usepackage{float}
\usepackage{color,xcolor}
\usepackage{graphicx,times}
\usepackage{cuted} \stripsep 8pt plus 2pt minus 2pt
\usepackage{epstopdf}
\usepackage{graphicx}
\usepackage{subfigure}
\usepackage{makecell}
\usepackage{multirow}
\renewcommand{\figurename}{Figure}
\usepackage[colorlinks,linkcolor=blue, urlcolor=blue, anchorcolor=blue, citecolor=blue]{hyperref}
\usepackage{longtable,booktabs}
\usepackage{fouriernc}
\usepackage{epsfig,graphics,graphicx}
\usepackage{color}
\usepackage{bm}
\usepackage{flushend}
\usepackage[sort&compress,square,numbers]{natbib}

\begin{document}
\title{A new tool for two-dimensional field-reversed configuration equilibrium study}

\author{
H. J. Ma$^{1,2,3}$, 
H. S. Xie$^{2,3*}$, 
B. H. Deng$^{2,3}$, 
Y. K. Bai$^{2,3}$, 
S. K. Cheng$^{2,3}$,
Y. Li$^{2,3}$, 
B. Chen$^{2,3}$, 
M. Tuszewski$^{4}$, 
H. Y. Zhao$^{2,3}$, 
and J.Y. Liu$^{1*}$
}

\address{
$^1$ Key Laboratory of Materials Modification by Laser, Ion and Electron Beams (Ministry of Education), School of Physics, Dalian University of Technology, Dalian 116024, People's Republic of China

$^2$ Hebei Key Laboratory of Compact Fusion, Langfang 065001, People’s Republic of China

$^3$ ENN Science and Technology Development Co., Ltd., Langfang 065001, People’s Republic of China

$^4$ ENN Consultant, Riverside, CA 92506, United States of America
}
\eads{*\mailto{xiehuasheng@enn.cn},  \mailto{jyliu@dlut.edu.cn}}

\begin{indented}
\item[\today]
\end{indented}

\begin{abstract}
A new tool (GSEQ-FRC) for solving two-dimensional (2D) equilibrium of field-reversed configuration (FRC) based on fixed boundary and free boundary conditions with external coils included is developed.
Benefiting from the two-parameter modified rigid rotor (MRR) radial equilibrium model and the numerical approaches presented by [Ma et al, Nucl. Fusion, 2021], GSEQ-FRC are used to study the equilibrium properties of FRC quantitatively and will be used for fast FRC equilibrium reconstruction.
In GSEQ-FRC, the FRC equilibrium can be  conveniently determined by two parameters, i.e., the ratio between thermal pressure and magnetic pressure at the seperatrix $\beta_s$, and the normalized scrape of layer (SOL) width $\delta_s$. 
Examples with fixed and free boundary conditions are given to demonstrate the capability of GSEQ-FRC in the equilibrium calculations.
This new tool is used to quantitatively study the factors affecting the shape of the FRC separatrix, revealing how the FRC changes from racetrack-like to ellipse-like.
\end{abstract}
%\keywords{FRC, Grad-Shafranov equation, MHD equilibrium, pressure profiles}
\submitto{\NF}
\maketitle
\ioptwocol

%===================================================================
\section{Introduction}
%===================================================================
Field-Reversed Configurations (FRC) is considered as one of the possible approaches to achieve fusion energy, either as  a magnetic confinement approach or as a target plasma for magnetized target fusion \cite{armstrong1981field,finn1982field,Tuszewski1988Field,steinhauer2011review, guo2015achieving}.
The plasma equilibrium is an essential element to understand the FRC properties \cite{ma2021two}, and is a foundation for studying various plasma phenomena such as magnetohydrodynamic (MHD) instabilities and plasma transport.

Understanding the interior properties of hot, dense FRC plasma is still a challenge.
In early FRC experiments, profiles information were deduced from measurements of the excluded flux array \cite{tuszewski1981excluded} and single-chord interferometry \cite{okada1989reduction}, 
and the shape of the separatrix can be estimated.
In recent FRC experiments, multi-point Thomson scattering  \cite{deng2012electron} and multi-chord interferometry \cite{deng2016high} provide more detailed FRC profile data. However,  non-perturbative internal magnetic field measurement is still difficult.
Therefore, theoretical/model equilibrium calculations are particularly important in the study of FRCs, and many equilibrium models such as Rigid-Rotor (RR) \cite{armstrong1981field}, two-point equilibrium (2PE) \cite{steinhauer2009equilibrium}, three-point equilibrium (3PE) \cite{steinhauer2014two}, symmetric (SYM) \cite{lee2020generalized} and Modified Rigid-Rotor (MRR ) \cite{ma2021two} are proposed.

The Grad-Shafranov (G-S) equation \cite{grad1958hydromagnetic, shafranov1966plasma} can be utilized to describe the traditional FRC equilibrium \cite{ma2021two}. And in cylindrical coordinate system $(r, z)$ with axial symmetry, it can be written as:
\begin{equation}
\Delta^*\psi\equiv r \frac{\partial}{\partial r}\left(\frac{1}{r} \frac{\partial \psi}{\partial r}\right)+\frac{\partial^{2} \psi}{\partial z^{2}}=-\mu_{0} r^2 P^{\prime}(\psi),  \label{eq:1}
\end{equation}
with
\begin{eqnarray}
J_{\theta}= r P^{\prime}(\psi), \label{eq:2} \\
B_z=\frac{1}{r} \frac{\partial{\psi}}{\partial{r}}; \quad B_r=-\frac{1}{r} \frac{\partial{\psi}}{\partial{z}},
\end{eqnarray}
where, $\psi=\Psi_p/2\pi=\int_{0}^{r} B_{z} r dr$ is the normalized poloidal magnetic flux, $J_{\theta}$ is toroidal current density of plasma, $P^{\prime}$ is the derivative of pressure with respect to $\psi$ and $\mu_{0}$ is the vacuum permeability.
In this work, zero toroidal magnetic field $B_{\theta} = 0$ is assumed.

Two-dimensional (2D) numerical equilibrium calculation of FRC maybe simpler than that for tokamak with omitted $B_{\theta}$. However, the requirement of including the plasma in the scrape of layer (SOL) region imposes some extra difficulties.
The existence of bifurcation solutions for the FRC equilibrium calculations was a major roadblock encountered by earlier FRC equilibrium research, and a 2D solution can be difficult to reach numerically  \cite{marder1970a, spencer1982free}.
Latter, the problem was overcome by using the total toroidal current ($I_{\theta}$) \cite{hewett1983free} in the computational region or the area of plasma surrounded by the separatrix ($S$)  \cite{suzuki1991effect, suzuki2000two} as the global constraints in equilibrium calculations. 
Based on $I_{\theta}$ and $S$, it is possible to solve FRC equilibrium with arbitrary elongation ratios.
In these traditional equilibrium calculations \cite{suzuki1991effect, suzuki2000two, steinhauer2014two, steinhauer2020anatomy, kako1983equilibrium, kanki1999numerical, gerhardt2006equilibrium}, iterative algorithms are used, the free model parameters usually are not physical parameters, therefore, post processing is required during and after the iteration to derive physical parameters for comparing with experimental data, or for merit assessment. The indirect algorithms are somehow inconvenient and inefficient in equilibrium studies, for example, in FRC equilibrium design. Sometimes more than two parameters are iterated, while as shown in reference \cite{ma2021two}, there are only two independent parameters needed to specify an FRC equilibrium in most models. The coupling of the model parameters could be the reason leading to bifurcated solutions.

In the recently proposed MRR model \cite{ma2021two}, FRC equilibrium can be determined by two  physical quantities $\beta_s$ and $\delta_s$, where $\beta_s$ is the ratio of the plasma  pressure to the magnetic pressure at the separatrix, and $\delta_s$ is the normalized SOL width.
The General Solver for EQuilibrium (GSEQ) code \cite{xie2019gseq} also utilizes the algorithm proposed in reference \cite{ma2021two} to solve 2D FRC equilibrium, with three steps: specifying the free parameters of the MRR model $\beta_s$ and $\delta_s$; solving other model parameters from 1D equilibrium; and solving 2D equilibrium.
Benefiting from this algorithm, the equilibrium calculation process almost becomes an algebraic equation solving process with high efficiency.
In addition, the FRC characteristics are readily obtained when the two model parameters $\beta_s$ and $\delta_s$ are determined. 
This greatly improve the efficiency to arrive at the desired equilibrium.
The convenience and efficiency of this GSEQ code can be very beneficial in the future for equilibrium fitting of experimental data.
One unique feature of the GSEQ-FRC  tool is that two key quantities of an FRC equilibrium, the maximum vacuum magnetic field $B_e$ (the detailed meaning of $B_e$ is discussed in \ref{Append:A}) and the separatrix radius $R_s$ are  scaling factors, which are not involved in the equilibrium solving process, and can be specified after the solution. This can significantly improve the computational efficiency in equilibrium design of FRCs.

This paper is organized as follows.
Section \ref{sec:level2} introduces the algorithms and procedures used in GSEQ-FRC.
In section \ref{sec:level3}, the 2D equilibrium with fixed boundary condition is presented.
Examples of free boundary equilibrium is shown in section \ref{sec:level4}.
In section \ref{sec:level5}, the factors affecting the shape of the FRC separatrix are discussed.
Summary and discussion are presented in section \ref{sec:level6}. 

%%%%%%%%%%%%%%%%%%%%%%%%%%%%%%%%%%%%%%%%%%%%%%%%%%%%%%%%%%%%%%%%%%%%%%
\section{\label{sec:level2} The algorithm in GSEQ-FRC}
The method for solving the free parameters of equilibrium models, and the initial result of 2D FRC equilibrium using the GSEQ code was presented in reference \cite{ma2021two}.
In the following, the 2D equilibrium solver is described in section \ref{sec:level2-1}. In Section \ref{sec:level2-2}, a step-by-step method for the GSEQ-FRC tool is presented.

\subsection{\label{sec:level2-1}Introduction to the G-S solver}
The GSEQ code, a plasma equilibrium solver with external coils, is developed for advanced study of plasma equilibrium. 
The accuracy of the result from GSEQ code has been demonstrated in \ref{Append:B}.
This code supports two distinctive `upper' ($r\to r_{\rm wall}$) boundary conditions: fixed boundary condition for metal wall as flux-conserver, and free boundary condition for quartz chamber.
Most of the algorithm for solving the G-S equation are well known, therefore, only a brief outline will be given.
In the flux-conserving metal wall, $\psi$ of  the `upper' boundary condition is calculated using Green's function \cite{gerhardt2006equilibrium}:
\begin{equation}
\psi_b(R, Z)=\iint G\left(R, Z ; R^{\prime}, Z^{\prime}\right) J_{\theta}\left(R^{\prime}, Z^{\prime}\right) d R^{\prime} d Z^{\prime}, \label{eq:4} 
\end{equation}
where 
\begin{equation}
G\left(R, Z ; R^{\prime}, Z^{\prime}\right) =\frac{\mu_{0}}{2 \pi} \frac{\sqrt{R R^{\prime}}}{k}\left[\left(2-k^{2}\right) K(k)-2 E(k)\right]   \label{eq:6} 
\end{equation}
is the free space Green's function, which gives the poloidal flux at ($R$,$Z$) from a unit coil current source at ($R^{\prime}$,$Z^{\prime}$),
$K(k)$ and $E(k)$ are the elliptic integrals of the first and the second kind respectively, and
\begin{equation}
k^{2} \equiv \frac{4 R R^{\prime}}{\left(R+R^{\prime}\right)^{2}+\left(Z-Z^{\prime}\right)^{2}}. \label{eq:5}
\end{equation}
In the free boundary condition case, a detailed procedure can be found in references \cite{johnson1979numerical, xie2019gseq}.
The `upper' boundary varies due to the changes of the plasma during the numerical iterations, and the G-S equation is solved inside the plasma-vacuum boundary. The $\psi_b$ on the edge of a computational domain is calculated as a combination of two parts:
\begin{equation}
\psi_{b}=\psi_{b}^{{coil} s}+\psi_{b}^{m}, \label{eq:7}
\end{equation}
where $\psi_b^{coils}$ represents the contribution from the currents in the external toroidal coils, and $\psi_{b}^{m}$ is the contributions from the plasma currents using the latest approximation to $J_{\theta}$  in the calculation proceedure. 
Both terms in equation \ref{eq:7} are computed using the Green's function for a toroidal current source.

\begin{figure}[h]
\includegraphics[width=8cm]{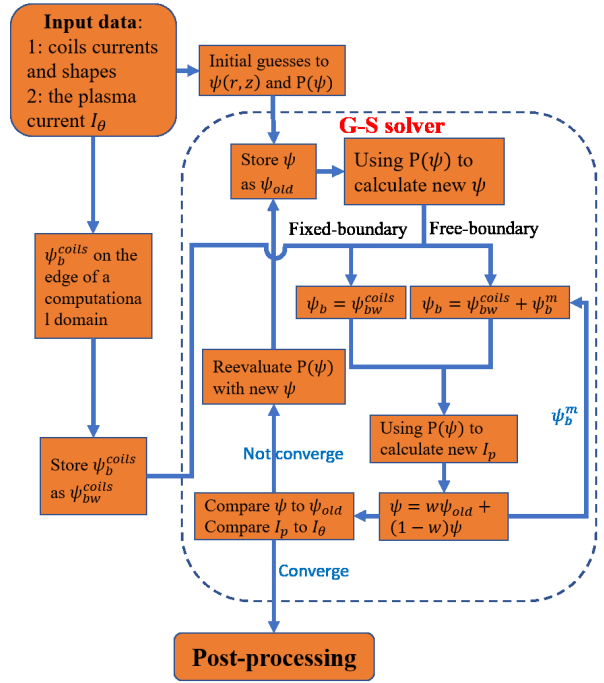}
\caption{Flow chart of the GSEQ code. }
\label{img_flow_chart_GSEQ}
\end{figure}

Next, we consider the computational procedures of the plasma subregion.
The plasma equilibrium in the GSEQ code is found by solving finite difference approximations to the G-S equation using the successive over relaxation (SOR) method\cite{hewett1983free, kako1983equilibria}.
For the FRC, the MHD equilibrium maybe more difficult with $r\to 0$ as the  `lower' boundary condition.
The numerical procedure is demonstrated in figure\ref{img_flow_chart_GSEQ} and comprises of three steps: 
In the first step, the information of the external coils including the current, the position and shapes, the computational domain and $I_{\theta}$ as the global constraints on the plasma equilibrium is given as the input data, and the guessed $\psi(r,z)$ is used as the initial $\psi$ distribution.
In the second step, the G-S equation solver is run. The new $\psi$ is updated using $P(\psi_{old})$ and the SOR method [$\psi=w \psi_{old}+(1-w) \psi$, $w\in(0,1)$ is the acceleration parameter] with fixed boundary or free boundary conditions respectively. 
The loop is based on the convergence of $\psi$ and $\psi_{old}$.
In the third step, the updating $\psi$ is output and post-processing when equation \ref{eq:8} is satisfied
\begin{equation}
d=max \lvert \psi(r,z)-\psi_{old} (r,z) \lvert <\varepsilon, \label{eq:8}
\end{equation}
where $\varepsilon$ is a specified small number, typically $10^{-9}$.

%%%%%
\subsection{\label{sec:level2-2}A step-by-step method for GSEQ-FRC}
\begin{figure}[h]
\includegraphics[width=8cm]{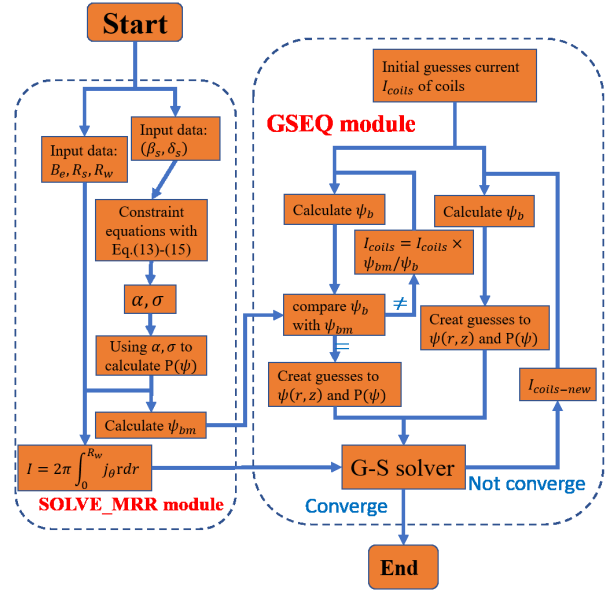}
\caption{The algorithm of GSEQ-FRC tool. }
\label{img_flow_chart_FRC_EFIT}
\end{figure}

In this subsection, GSEQ-FRC tool is presented, which combines the method for solving the free parameters of MRR model \cite{ma2021two} and the GSEQ code.
The flowchart of the GSEQ-FRC tool is illustrated in figure \ref{img_flow_chart_FRC_EFIT}.
As described in reference \cite{ma2021two}, the MRR-1 model is:
\begin{equation}
P(\psi)=\frac{B_{e}^{2}}{2 \mu_{0}} \beta_{s} \cdot \exp \left(-\alpha \frac{ \psi}{B_{e} R_{s}^{2}} \right) \cdot\left[\sigma \Big(\frac{ \psi}{B_{e} R_{s}^{2}} \Big)^q+1\right]^{n},\label{eq:9}
\end{equation}
with $q=1$, $n=2$ as default, $\alpha$ and $\sigma$ are the free parameters.
The free parameters in equation \ref{eq:9} are solved from the following normalized constraint equations \cite{ma2021two}:
\begin{eqnarray}
&\delta_{s}=\left.-\frac{p}{\frac{d p}{d r}}\right |_{r=R_s}=\left.-\frac{p(\psi)}{\frac{d p(\psi)}{d \psi} \cdot \frac{d \psi}{d r}}\right |_{r=R_s},  \label{eq:10}\\
&p\left(\psi_{m}\right)-1=0,  \label{eq:11}\\
&\int_{0}^{\psi_{m}} \frac{d \psi}{\sqrt{\left[1-p\left(\psi_{m}\right)\right]}}+\frac{1}{4}=0, \label{eq:12}
\end{eqnarray}
where $\delta_s$ is used to characterize the SOL width, and $\psi_m$ is the trapped flux.
The dimensionless forms are used for the radius  $r/R_s\to r$, the flux $\psi / (B_e R_{s}^2) \to \psi$, and the pressure $P/P_{m}\to p$, where $P_m=B_e^2/ 2\mu_0$.
The detailed  of equations  \ref{eq:10}-\ref{eq:12} has been shown in reference \cite{ma2021two}.
Using equations \ref{eq:10}-\ref{eq:12}, the FRC equilibrium is uniquely determined from $\beta_s$ and $\delta_s$.

In GSEQ-FRC tool, the important quantities $B_e$, $R_s$ and $R_w$ (the chamber radius) are not involved in the equilibrium solving process,
which is because the $I_{1D}=2\pi \int_{0}^{R_w}j_{\theta}r dr$ as a global constraint instead of $I_{\theta}$ in the 2D equilibrium solving of GSEQ-FRC.
The following steps are used: Specifying two parameters $\beta_{s}$ and $\delta_s$ to obtain the normalized 1D equilibrium profiles.
With the target values of $B_e$, $R_s$ and $R_{w}$, the poloidal flux at the chamber wall $\psi_{bm}$ is determined, which is used to calculate the external coil current of the fixed boundary condition.
Meanwhile, $I_{1D}$ is also calculated to replace $I_{\theta}$ in GSEQ module.
The GSEQ module is run and the 2D equilibrium is uniquely determined when the $\psi$ value converges.
Subsequently, the 2D equilibrium properties are obtained.

%%%%%%%%%%%%%%%%%%%%%%%%%%%%%%%%%%%%%%%%%%%%%%%%%%%%%%%%%%%%%%%%%%%%%%
\section{\label{sec:level3} 2D equilibrium with fixed boundary condition}
In this section, we apply the GSEQ-FRC to study some typical FRC equilibria.

\subsection{2D equilibrium with fixed boundary condition}
\begin{figure}[ht]
\includegraphics[width=8cm]{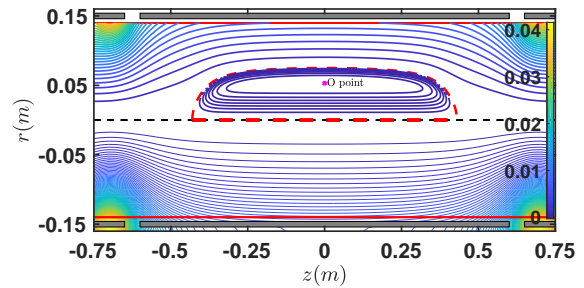}
\caption{The schematic of the FRC machine, where the gray boxes are the external coils, and the solid red line is the wall. The upper half shows the magnetic fluxes with the FRC included, and the lower half shows the magnetic fluxes of vacuum field.}
\label{img_FRC_shape}
\end{figure}

A simple FRC machine with three coils is shown in figure \ref{img_FRC_shape}.
The central confinement region has  0.28 m inner diameter and 1.5 m length.
The settings of the external coils are presented in \Tref{tab-1}.
In this subsection, the wall is assumed to be ideal conductor (here we mean that the flux $\psi$ at wall is taken as the vacuum flux $\psi^{coil}$ from the outside coils), so the equilibrium can be solved with the procedure for fixed boundary case as discussed above. 

The target values of $B_e$, $R_s$ and $R_w$ are not involved in the equilibrium solving when using GSEQ-FRC for the design of the 2D equilibrium.
This new tool eliminates the requirement for numerous calculations to match with the design values.
In the process of GSEQ-FRC solving 2D equilibrium, the $I_{1D}$ is used as the global constraint instead of $I_{\theta}$, which is the first time to link 1D equilibrium characteristics with 2D.
In the following, $B_e=1 T$ and $R_s=7.5 cm$ are taken as the target values.

\begin{table}[htbp]
\caption{The external coils parameters.}\label{tab-1}
\begin{indented}
\item[]\begin{tabular}{@{}llllll}
\br
Coil No. &Z (m)&R (m)& Width (m)&Hight (m)\\
\mr
1&-0.7&0.15&0.1&0.006\\
2&0.0&0.15&1.2&0.006\\
3&0.7&0.15&0.1&0.006\\
\br
\end{tabular}
\end{indented}
\end{table}

\begin{figure}[ht]
\includegraphics[width=8cm]{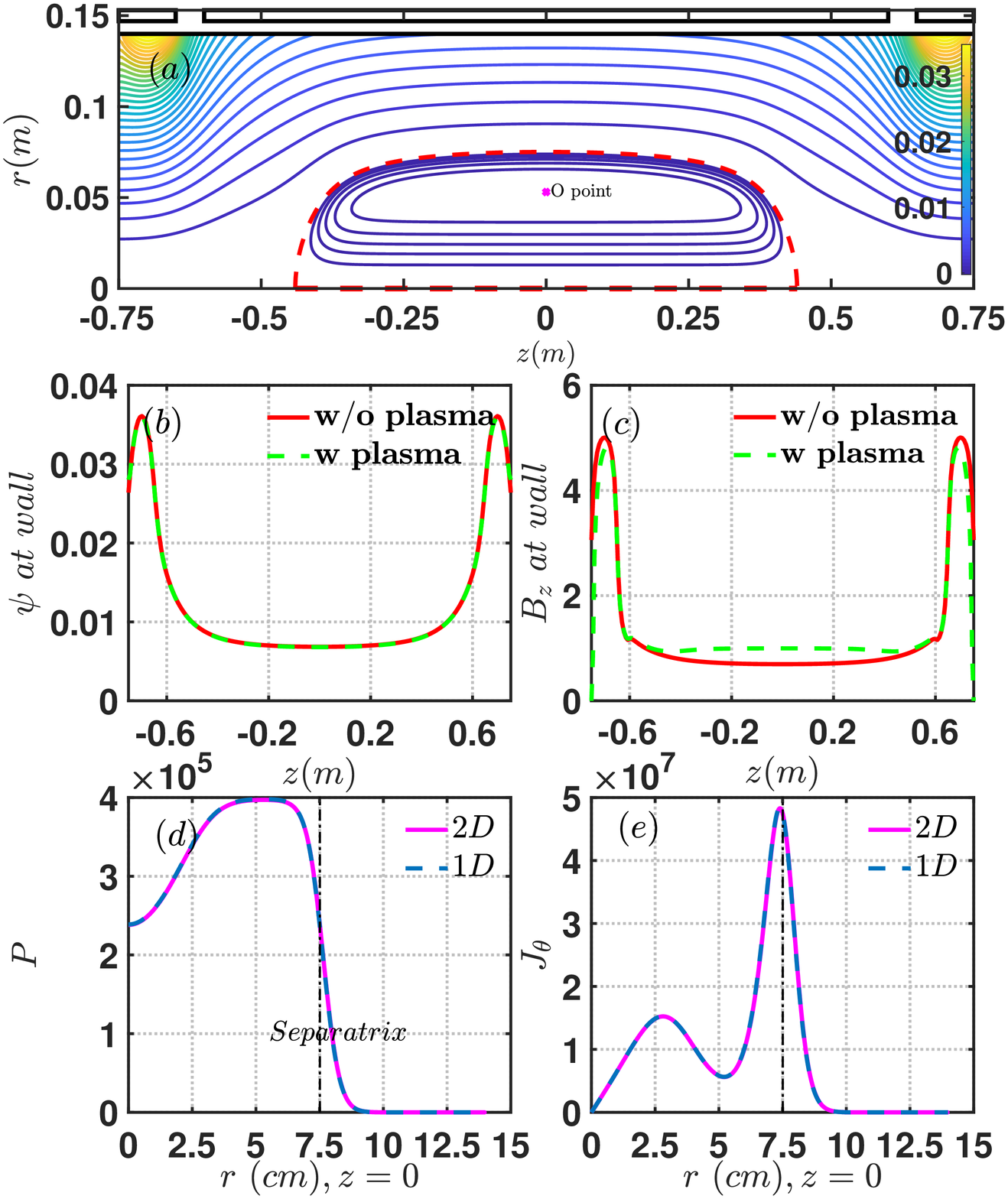}
\caption{The 2D fixed boundary condition equilibrium with the $B_e=1 T$, $R_s=0.075m$, $\beta_s=0.6$ and $\delta_s=0.4\delta_{s}^{RR}$, where $\delta_{s}^{RR}$ is the $\delta_s$ of RR model. (a) is the contour of magnetic fluxes of 2D equilibrium from GSEQ-FRC. (b), (c) are the profiles of $\psi(z)$ and $B_z(z)$ respectively, where the red solid line is the profile in vacuum and the green dash line is the profile of final convergent results. (d), (e) are the equilibrium pressure and current density profiles of 1D and the 2D in midplane respectively. The external coil currents are all $6.45 \times 10^{5}$ (A) for each coils.}
\label{img_eq_with_fix}
\end{figure}

The 2D equilibrium of FRC using fixed boundary condition is demonstrated in figure \ref{img_eq_with_fix}.
The contour of the 2D equilibrium is shown in figure \ref{img_eq_with_fix} (a).
The axial distribution of $\psi$ and magnetic field $B_z$ at the wall are shown in figures \ref{img_eq_with_fix} (b) and (c), respectively, with external currents of $6.45 \times 10^{5}$ (A) for each coils. 
The 1D equilibrium is the input data in the process of solving 2D equilibrium using GSEQ-FRC tool, so the comparison of pressure and current density profiles from solve\_MRR module (1D) and the midplane profiles (2D) of 2D equilibrium are shown in figures \ref{img_eq_with_fix} (d) and (e) respectively as a confidence check, which is satisfactorily.

\subsection{Equilibrium with complex fixed boundary condition}
\begin{figure}[ht]
\includegraphics[width=8cm]{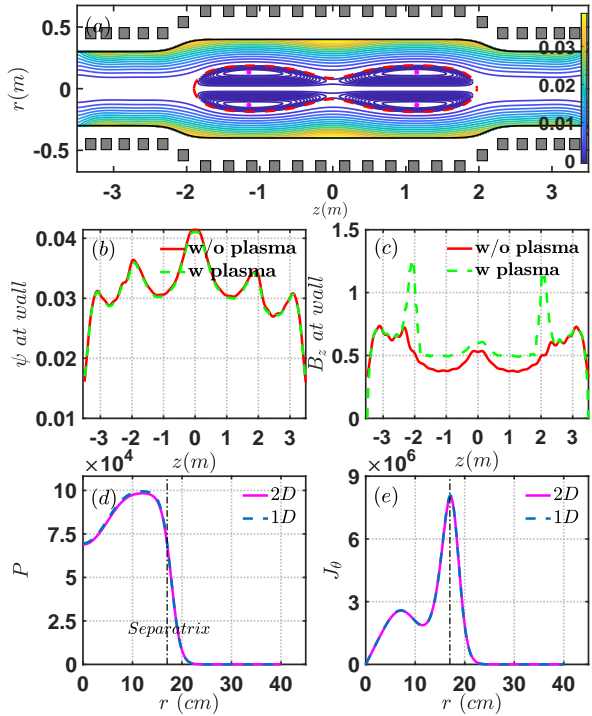}
\caption{The complexed 2D equilibrium with non-regular boundary and asymmetric coil currents. (a) is the contour of magnetic fluxes of 2D equilibrium, (b), (c) is the profiles of $\psi$ and $B_z$ with $z$, (d), (e) is the pressure and current density profiles of 1D and O-point of left-side FRC respectively.}
\label{img_fix_merge}
\end{figure}

In this subsection, the complex equilibrium with irregular boundary and asymmetric coil currents is reconstructed based on the FIX device \cite{yambe2009equilibrium}, which is shown in figure \ref{img_fix_merge}.
The equilibrium of the two mergeing FRCs is demonstrated in figure \ref{img_fix_merge} (a).
It should be noticed that the currents of the external coils are asymmetric, which can be seen from $\psi$ and $B_z$ profile at the walls of figures \ref{img_fix_merge} (b) and (c), respectively.
The complex equilibrium is solved using the $I_{1D}$ from MRR\_module as a constraint with the total current at the O-point of left-side FRC. 
The pressure and current density profiles of 1D equilibrium are consistent with the poloidal profiles of 2D equilibrium at the O-point, which is presented in figures \ref{img_fix_merge} (d) and (e).

%%%%%%%%%%%%%%%%%%%%%%%%%%%%%%%%%%%%%%%%%%%%%%%%%%%%%%%%%%%%%%%%%%%%%%
\section{\label{sec:level4} 2D equilibrium with free boundary condition}
\begin{figure}[ht]
\includegraphics[width=8cm]{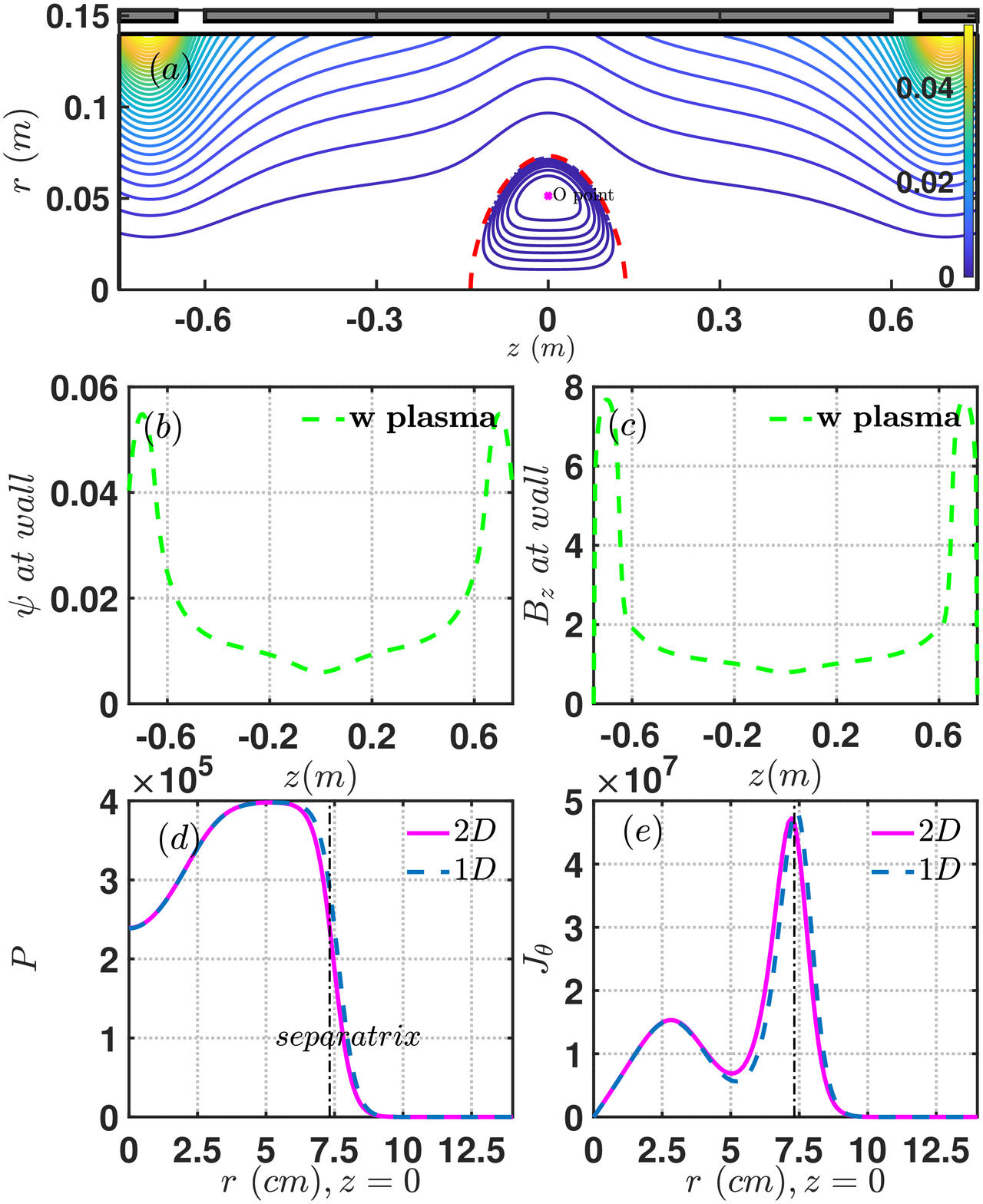}
\caption{The 2D free boundary condition equilibrium with the setting of $B_e=1 T$, $R_s=0.075m$, $\beta_s=0.6$ and $\delta_s=0.4\delta_{s}^{RR}$. (a) is the contour of magnetic fluxes of 2D equilibrium, (b), (c) is the final convergent profiles of $\psi$ and $B_z$ with $z$, (d), (e) is the pressure and current density profiles of 1D and midplane of 2D equilibrium respectively.}
\label{img_eq_with_free}
\end{figure}

In this section, the 2D equilibrium with free boundary condition is calculated using GSEQ-FRC.
As shown in figure \ref{img_eq_with_free}, the 2D free boundary condition equilibrium is presented with the same settings as those in the fixed boundary condition case. 
Comparing figure \ref{img_eq_with_free} (a) with figure \ref{img_eq_with_fix} (a), it is found that the 2D equilibrium of free boundary and fixed boundary of FRC are significantly different, which is reasonable because that the boundary fluxes are very different as can be seen from figure  \ref{img_eq_with_free} (b) and figure \ref{img_eq_with_fix} (b).
In figure \ref{img_eq_with_free} (c), it is found that $B_z$ at the midplane is not the preset $B_z=1T$, which is explained in \ref{Append:A}.
Also, the pressure and current density profiles of 1D equilibrium match well with the midplane profiles of 2D equilibrium, which is illustrated in figures \ref{img_eq_with_free} (d) and (e).

\begin{figure}[ht]
\includegraphics[width=8cm]{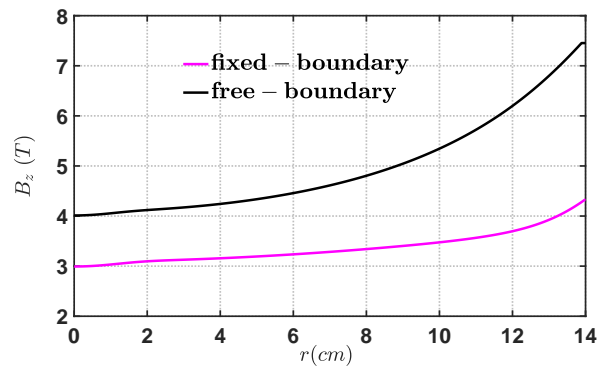}
\caption{The $B_z$ radial profiles of fixed (magenta line) and free (black line) boundary condition equilibrium near the left boundary ($z=-0.72 m$).}
\label{img_Bz_fixfree}
\end{figure}
The $B_z$ radial profiles near the left boundary ($z=-0.72 m$) from fixed and free boundary are shown in figure \ref{img_Bz_fixfree}.
It is found from figure \ref{img_Bz_fixfree} that $B_z$ from free boundary equilibrium is larger, i.e., the magnetic pressure is greater, for which the plasma is more compressed axially, so that the elongation ratio of FRC from free boundary condition equilibrium is smaller than that from fixed boundary equilibrium.

We have found \cite{ma2021two} that four independent parameters $(\beta_s, \delta_s, B_e, R_s)$ can capture most of the FRC 1D equilibrium profile features. Whereas, five independent parameters $(\beta_s, \delta_s, B_e, R_s, l_s)$ can probably roughly describe a 2D FRC equilibrium. In the present work, we can control the first four parameters $(\beta_s, \delta_s, B_e, R_s)$  well. However, how to control the FRC length $l_s$ is still an open question, especially in the free boundary condition case.

For the case of free boundary condition, the method used in the present version of GSEQ-FRC is similar to the one that solve the free boundary equilibrium of traditional tokamak \cite{johnson1979numerical}.
One difference between FRC and traditional tokamak is that the plasma exists in the region outside the separatrix of the FRC, i.e., the coupling between FRC plasma and coil is stronger.
Due to this, the different computation positions of the left and right boundaries will also affect the results.
We also notice some previous studies of 2D FRC equilibria \cite{spencer1982free,kako1983equilibria,fuentes1995ecmc,kanki1999numerical},  which are relevant to free boundary condition equilibria. However, they are not the same as our treatment or the standard treatment in tokamak\cite{johnson1979numerical} of the free boundary equilibrium. Therefore, the 2D free boundary condition equilibrium of FRC requires further investigation.

%%%%%%%%%%%%%%%%%%%%%%%%%%%%%%%%%%%%%%%%%%%%%%%%%%%%%%%%%%%%%%%%%%%%%%
\section{\label{sec:level5} The factors affecting the shape of FRC separatrix}
In early 2D numerical equilibrium studies, the racetrack-like separatrix is often obtained. However, both racetrack-like and ellipse-like separatrices were shown in early experiments.
Reference \cite{spencer1985experimental} investigated the factors affecting the shape of the separatrix based on MHD equilibrium, and \textcolor{black}{concluded that steeper $P(\psi)$ can cause more ellipse-like separatrix}.  However, references \cite{suzuki1999analysis, suzuki2000two} also studied this problem but obtained an opposite conclusion, i.e., the shape of the separatrix becomes racetrack-like as $P(\psi)$ becomes steeper at the separatrix.
If we check how references \cite{spencer1985experimental, suzuki1999analysis, suzuki2000two} obtained their conclusions, we find those conclusions are only qualitative. For example, the method that caused the steep $P(\psi)$ was not mentioned clearly in reference \cite{spencer1985experimental}. Whereas, the parameter $\gamma$ increases with the other parameters also be changed to achieve the ellipse-like shape in references \cite{suzuki1999analysis, suzuki2000two} (cf. the model parameters of subplots (a) in figure 5 and 6 from reference \cite{suzuki2000two}).

The factors affecting the shape of the separatrix are investigated quantitatively in this subsection to resolve the above confusion.
We use the following equation to describe the shape of the separatrix, which is defined in reference \cite{ohkuma2010separatrix} with
\begin{equation}
\frac{r^{2}}{a^{2}}+\frac{|z|^{m}}{b^{m}}=1, \label{eq:15}
\end{equation}
where $a$ is the radius of the separatrix at the midplane, $b$ is the half-length of the separatrix, and $m$ is the shape index.
The larger the parameter $m$ is the more racetrack-like shape of separatrix.
$m=2$ is the elliptical shape in equation \ref{eq:15}.

\begin{figure}[ht]
\includegraphics[width=8cm]{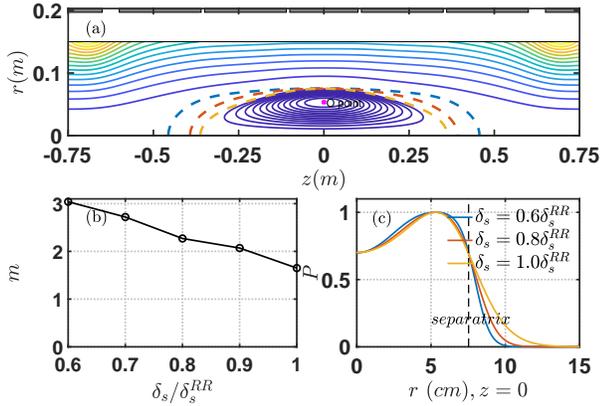}
\caption{The changes of 2D equilibrium for different $\delta_s$ with $\beta_s=0.6$. (a) is the shapes of the separatrix, (b) is the changes of $m$ with $\delta_s$, (c) is the normalized pressure profiles.  The chamber has a length of 1.5 m and a diameter of 0.3 m, with the external coils at 0.2 m. The other values are set to $B_e=1 T$, $R_s=0.075m$. }
\label{img_shape_deltas}
\end{figure}

The fixed boundary condition equilibrium are shown in figure \ref{img_shape_deltas} with varying $\delta_s$ but fixed $\beta_s$. 
As $\delta_s$ decreases, the shape of the separatrix presented in figure \ref{img_shape_deltas} (a) changes from ellipse-like to racetrack-like, and the parameter $m$ that best fits the shape of the separatrix with equation\ref{eq:15} increases as shown in figure \ref{img_shape_deltas} (b). 
In this subsection, we use $-dp/dr$ to represent the pressure gradient.
The pressure gradient near the separatrix is lager with the shape of separatrix becoming racetrack-like, which is demonstrated in figure \ref{img_shape_deltas} (c).

\begin{figure}[ht]
\includegraphics[width=8cm]{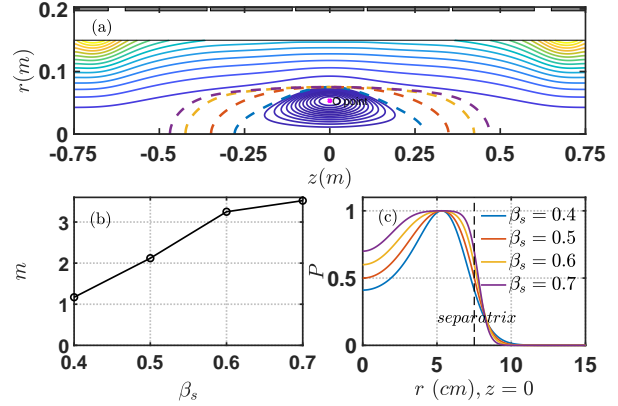}
\caption{The changes of 2D equilibrium for different $\beta_s$ with $\delta_s=0.15$. (a) is the shapes of the separatrix, (b) is the changes of $m$ with $\beta_s$, (c) is the normalized pressure profiles. The other values are set to $B_e=1 T$, $R_s=0.075m$. }
\label{img_shape_betas}
\end{figure}

Figure \ref{img_shape_betas} illustrates the changes of fixed boundary condition equilibrium for different $\beta_s$ with fixed $\delta_s$. 
When $\beta_s$ increases, the shape of the separatrix changes from ellipse-like to racetrack-like, and the parameter $m$ also increases, which are presented in figure \ref{img_shape_betas} (a) and (b), respectively.
Once again, as shown in figure \ref{img_shape_betas} (c), the pressure gradient becomes larger with the shape of separatrix becomes racetrack-like.
From figures \ref{img_shape_deltas} and \ref{img_shape_betas}, we find that decreasing $\delta_s$ or increasing $\beta_s$ make the pressure gradient near the separatrix lager, and the shape of the separatrix becomes more racetrack-like.

Next, we do an analysis of the results in references \cite{spencer1985experimental, suzuki1999analysis, suzuki2000two}, where the derivative of the pressure with respect to $\psi$ is
\begin{equation}
\frac{d P(\psi)}{d \psi}=\left\{\begin{array}{ll}
-c(1+\epsilon \psi) &\quad(\psi \leqslant 0) \\
-c e^{-\gamma \psi} &\quad(\psi>0)
\end{array}\right. ,\label{eq:13}
\end{equation}
where $c$, $\epsilon$ and $\gamma$ are constants. 
$\delta_s$ from equation\ref{eq:13} is expressed as 
\begin{equation}
\delta_s=\frac{1}{\gamma B_e R_s^2 \sqrt{1-\beta_s}}. \label{eq:16}
\end{equation}
From equation \ref{eq:16}, it is found that the increase in  $\gamma$ causes a decrease in $\delta_s$, thus a steeper pressure near the separatrix, and the shape of separatrix changes from ellipse-like to racetrack-like.
Equation \ref{eq:16} reasonably explains the conclusions in references \cite{suzuki1999analysis, suzuki2000two}.
The change of $\beta_s$ can also cause a change of $\delta_s$, which is presented in equation \ref{eq:16}.
However, when $\delta_s$ is fixed, the pressure gradient increases with $\beta_s$, and the shape of the separatrix becomes  racetrack-like, which would be similar to figure \ref{img_shape_betas}. Thus, the results in references \cite{suzuki1999analysis, suzuki2000two} are consistent with ours. However, our results are more clearly and conclusive due to that we can control the pressure gradient parameter explicitly and quantitatively by using $\delta_s$ and $\beta_s$.
A simple physics picture is that racetrack-like equilibrium has larger axial gradient than the ellipse-like equilibrium, and therefore corresponds to higher radial gradient.

%%%%%%%%%%%%%%%%%%%%%%%%%%%%%%%%%%%%%%%%%%%%%%%%%%%%%%%%%%%%%%%%%%%%%%
\section{\label{sec:level6} Summary and conclusion}
The GSEQ-FRC tool, a 2D FRC G-S equilibrium simulation tool has been developed and is applied to FRC equilibrium design.
Unlike the conventional method, in GSEQ-FRC, the equilibrium are solved with  two parameters using the physical properties of FRC.
Several examples with both fixed boundary and free boundary conditions have demonstrated its powerful capabilities.
The properties of the FRC equilibrium with fixed and free boundaries are investigated systematically.
Furthermore, the factors affecting the shape of the FRC separatrix are discussed quantitatively, and it is found that steeper pressure at the separatrix causes the shape of separatrix to change from ellipse-like to racetrack-like. 
The GSEQ-FRC tool has advantages in equilibrium calculations, which can also be extended to FRC equilibrium reconstruction. The application of the GSEQ-FRC tools to specific experiments data will be shown in the future publications.

\ack
This work is supported by the China central government guides the development of local science and technology funding No.206Z4501G and the compact fusion project of the ENN group.
One of the author HSX would also like to thank T. Takahashi in Gunma University for the help in benchmarking the GSEQ code.

\appendix{}
%%%%%%%%%%%%%%%%%%%%%%%%%%%%%%%%%%%%%%%%%%%%%%%%%%%%%%%%%%%%%%%%%%%%%%
\section{\label{Append:A} The accuracy of the quasi-1D equilibrium}
From the equation (B.4) in reference \cite{ma2021two}, the general 1D force balance equation from ${\bm  J}\times {\bm B}=\nabla P$ is
\begin{equation}
P(r)+\frac{B_{z}^{2}(r)}{2 \mu_{0}}=\frac{B_{e}^{2}}{2 \mu_{0}}+\int \frac{B_{z}}{\mu_{0}}\left(\frac{\partial B_{r}}{\partial z}\right) d r. \label{appendA-2}
\end{equation}
The second term on the right side is the magnetic field curvature effect at the midplane.
In equation \ref{appendA-2}, we define $B_e\equiv\sqrt{2\mu_0 P_m}$, where $P_m$ is the maximal pressure, i.e., the pressure at the O-point. 
The quasi-1D equilibrium $P+\frac{B_{z}^{2}}{2 \mu_{0}}=\frac{B_{e}^{2}}{2 \mu_{0}}=P_m$ ignoring the curvature effect can describe the elongated cylindrical FRC well, where $B_e$ can be also defined as the vacuum field $B_z(r\to \infty)$ or $B_z(r\to wall)$ where $P\to0$. The quasi-1D equilibrium approximation is commonly used in both theoretical and experimental studies.

\begin{figure}[ht]
\includegraphics[width=8cm]{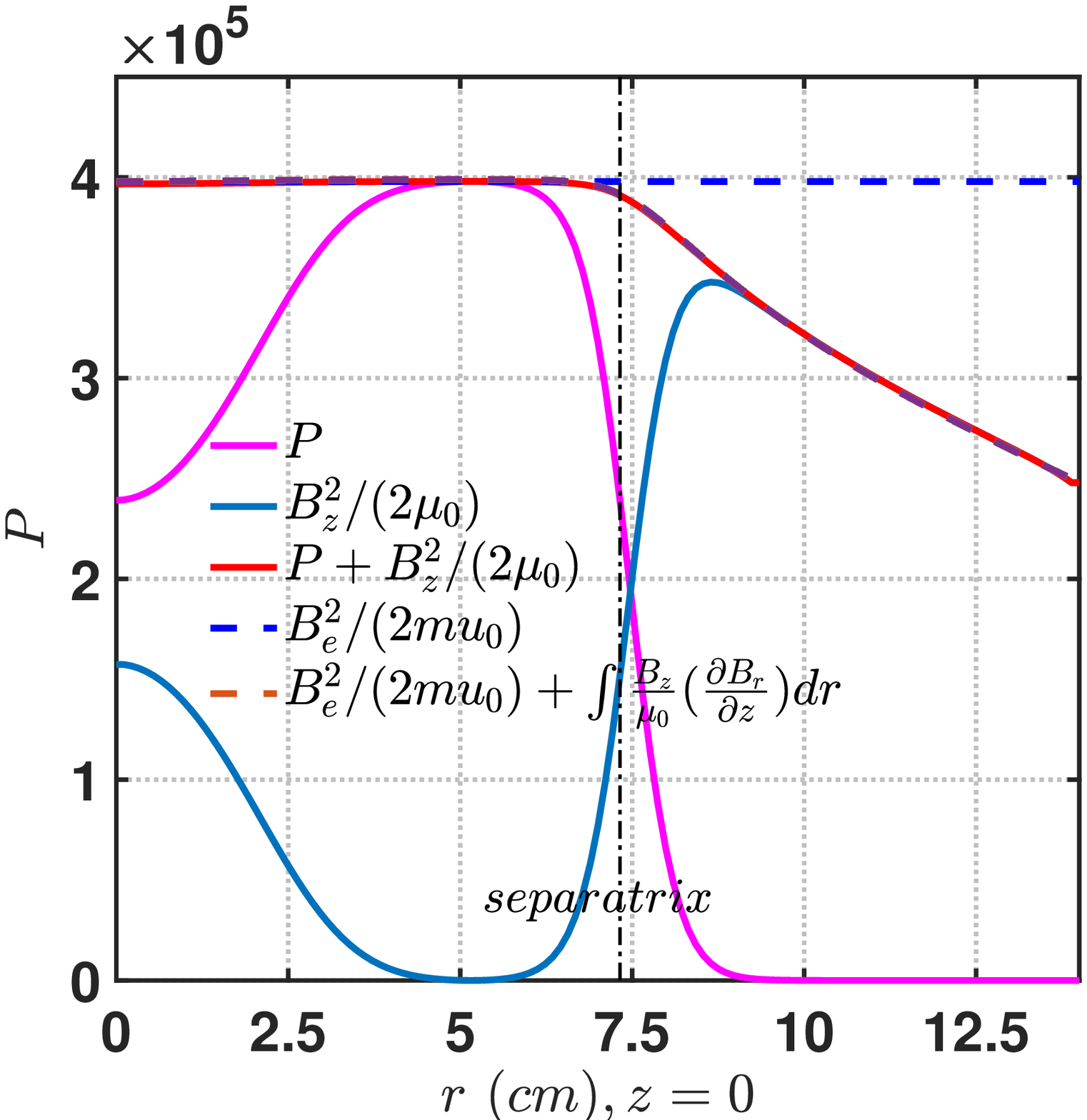}
\caption{The radial profiles of the free boundary equilibrium at the midplane in figure \ref{img_eq_with_free}.}
\label{img_eq-1D}
\end{figure}

Figure \ref{img_eq-1D} shows the radial profiles of the free boundary equilibrium at the midplane in figure \ref{img_eq_with_free}.
From the red solid line and blue dashed line in figure \ref{img_eq-1D}, it is found that the quasi-1D equilibrium matches well inside the separatrix, but the deviation increases outside the separatrix.
However, the left and right sides of equation \ref{appendA-2} balanced well, which is demonstrated by the red solid line and the brown dashed line.
Although it is found from figure \ref{img_eq-1D} that the magnetic field curvature effect is significant outside the separatrix in the 2D free boundary condition equilibrium, the 2D results of GSEQ for $P(r)$ and $J_\theta(r)$ still match well with the 1D results, which is shown in figure \ref{img_eq_with_free} (d) and (e). This implies that the solution approach used in GSEQ-FRC based on MRR model is robust for even the midplane magnetic field curvature effect is non-ignorable.

%%%%%%%%%%%%%%%%%%%%%%%%%%%%%%%%%%%%%%%%%%%%%%%%%%%%%%%%%%%%%%%%%%%%%%
\section{\label{Append:B} The benchmark of the GSEQ code}
The GSEQ code is benchmarked with  grass\_ft \cite{takahashi2004losses}. 
The  grass\_ft is a FRC G-S equation solver with fixed boundary condition.
In grass\_ft code, the `upper' boundary $\psi(r_w,z)$ is set with $r_w =const$:
\begin{equation}
\psi=\left\{\begin{array}{l}
\psi_{w} ; \quad 0 \leqslant z \leqslant z_{c} \\
\frac{r_{\psi}+1}{2} \psi_{w}+\frac{1-r_{\psi}}{2} \psi_{w} \cos \left(\frac{z-z_{c}}{z_{m i r}-z_{c}} \pi\right) ; z>z_{c}
\end{array}\right. . \label{append-1}
\end{equation}
Here,  $z_{mir}$ is the axial length from the midplane to the mirror end, $z_c$ is the axial position at which the mirror field critically influences, and $r_{\psi}$ is a control parameter for the mirror ratio.
The pressure profile with $\psi$ is 
\begin{equation}
p(\psi)=\left\{\begin{array}{l}
p_{s}+\frac{p_{s}}{\psi_{w}} \ln \left(\frac{p_{w}}{p_{s}}\right) \psi \\
+\frac{1}{2} \frac{p_{s}}{\psi_{w}^{2}}\left[\ln \left(\frac{p_{w}}{p_{s}}\right)^{2}\right] \psi^{2} \quad(\psi \geq 0) \\
p_{s}\left(\frac{p_{w}}{p_{s}}\right)^{\psi / \psi_{w}} \quad(\psi<0) 
\end{array}\right. , \label{append-2}
\end{equation}
where $p_s$, $\psi_w$, $p_w$ are coefficients. In equation \ref{append-2}, the normalized parameter of $p(\psi)$ is $\psi_w^2/2\mu_0 r_w^4$. Note that in grass\_ft code, $\psi>0$ is inside the separatrix as shown in equation \ref{append-2}, which is opposite to the default definition in GSEQ code. We thus change the GSEQ code to do this benchmark.

\begin{figure}[ht]
\includegraphics[width=8cm]{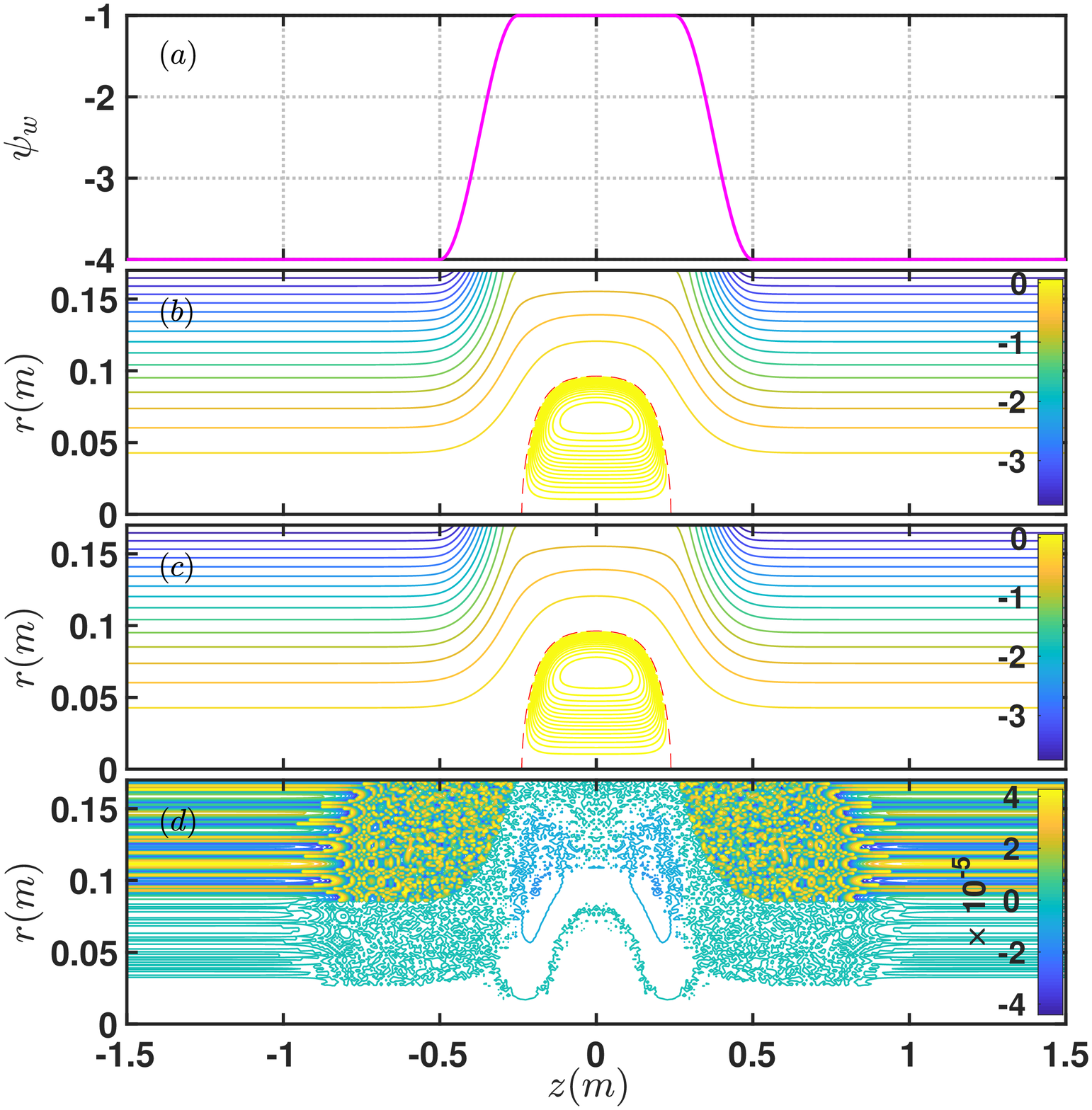}
\caption{The comparison of the results of GSEQ and grass\_ft code. (a) is the ``upper" boundary with $r_{\psi}$=-4, $\psi_w=$-1, $z_c$=0.25 and $z_{mir}$=0.5, (b) is the contour of $\psi(r,z)$ from GSEQ code, (c) is the contour of $\psi(r,z)$ from grass\_ft code, and (d) is the difference of $\psi$ between GSEQ and grass\_ft code. The other values are set to $R_{w}=0.17m$ and $Z_{w}=1.5m$.}
\label{img_contour}
\end{figure}

The results calculated by GSEQ and grass\_ft code are shown in figure \ref{img_contour}, where the free parameters in equation \ref{append-2} are set with $p_s$=5.5, $\psi_w$=1 and $p_w=1\times10^{-5}$.
Figure \ref{img_contour} (a) is the specific fixed boundary set by equation \ref{append-1}.
Figure \ref{img_contour} (b) and (c) shows the contour of $\psi(r,z)$ from GSEQ and grass\_ft equilibrium codes with the same parameters respectively.
Figure \ref{img_contour}(d) illustrates that the difference of $\psi(r, z)$ between GSEQ and grass\_ft code is less than $4\times10^{-5}$, which means that the GSEQ code agrees well with grass\_ft code.

\begin{thebibliography}{26}
\makeatletter
\renewcommand{\@biblabel}[1]{[#1]}
\makeatother
\bibitem{armstrong1981field}Armstrong W. T. $\textit{et al}$. 1981, Field-reversed experiments (FRX) on compact toroids. $\textit{Phys. Fluids}$, 24, 2068
\bibitem{finn1982field}Finn J. M. $\textit{et al}$. 1982, Field-reversed configurations with a component of energetic particles. $\textit{Nucl. Fusion}$, 22, 1433
\bibitem{Tuszewski1988Field}Tuszewski M. 1988, Field reversed configurations. $\textit{Nucl. Fusion}$, 28, 2033
\bibitem{steinhauer2011review}Steinhauer L. C. 2011, Review of field-reversed configurations. $\textit{Phys. Plasmas}$, 18, 070501
\bibitem{guo2015achieving}Guo H. Y.  $\textit{et al}$. 2015, Achieving a long-lived high-beta plasma state by energetic beam injection. $\textit{Nat. Commun.}$, 6, 6897
\bibitem{ma2021two}Ma H. J. $\textit{et al}$. 2021, Two-parameter modified rigid rotor radial equilibrium model for field-reversed configurations. $\textit{Nucl. Fusion}$, 61, 036046
\bibitem{tuszewski1981excluded}Tuszewski M. 1981, Excluded flux analysis of a field-reversed plasma. $\textit{Phys. Fluids}$, 24, 2126
\bibitem{okada1989reduction}Okada S. $\textit{et al}$. 1989, Reduction of the density profile of a field-reversed configuration plasma from detailed interferometric measurements. $\textit{Jpn. J. Appl. Phys.}$, 65, 4625
\bibitem{deng2012electron}Deng B. H. $\textit{et al}$. 2012, Electron density and temperature profile diagnostics for C-2 field reversed configuration plasmas. $\textit{Rev. Sci. Instrum.}$, 83, 10E339
\bibitem{deng2016high}Deng B. H. $\textit{et al}$. 2016, High sensitivity far infrared laser diagnostics for the C-2U advanced beam-driven field-reversed configuration plasmas. $\textit{Rev. Sci. Instrum.}$, 87, 11E125
\bibitem{steinhauer2009equilibrium}Steinhauer L. C. $\textit{et al}$. 2009, Equilibrium paradigm for field-reversed configurations and application to experiments. $\textit{Phys. Plasmas}$, 16, 072501
\bibitem{steinhauer2014two}Steinhauer L. $\textit{et al}$. 2014, Two-dimensional interpreter for field-reversed configurations. $\textit{Phys. Plasmas}$, 21, 082516
\bibitem{lee2020generalized}Lee K. Y. 2020, Generalized radial profile of field-reversed configurations based on symmetrical properties. $\textit{Nucl. Fusion}$, 60, 046010
\bibitem{grad1958hydromagnetic}Grad H. $\textit{et al}$. 1958, Hydromagnetic equilibrium and Force-Free Fields. $\textit{Journal of Nuclear Energy}$, 7, 284
\bibitem{shafranov1966plasma}Shafranov V. D. $\textit{et al}$. 1966, Plasma equilibrium in a magnetic field. $\textit{Reviews of Plasma Physics}$, 2, 130
\bibitem{marder1970a}Marder B. $\textit{et al}$. 1970, A vifurcation problem in E-layer equilibrium. $\textit{plasma physics}$, 12, 435
\bibitem{spencer1982free}Spencer R. L. $\textit{et al}$. 1982, Free boundary field-reversed configuration (FRC) equilibrium in a conducting cylinder. $\textit{Phys. Fluids}$, 25, 1365
\bibitem{hewett1983free}Hewett D. W. $\textit{et al}$. 1983, Two-dimensional equilibrium of field-reversed configurations in a perfectly conducting cylindrical shell. $\textit{Phys. Fluids}$, 26, 1299
\bibitem{suzuki1991effect}Suzuki K. 1991, Effect of the Mirror Field on the Averaged $\beta$ Value in Field Reversed Configuration. $\textit{Jpn. J. Appl. Phys.}$, 60, 3186
\bibitem{suzuki2000two}Suzuki Y. $\textit{et al}$. 2000, Two-dimensional numerical equilibrium of field-reversed configuration in the strong mirror field. $\textit{Phys. Plasmas}$, 7, 4062
\bibitem{steinhauer2020anatomy}Steinhauer L. C. $\textit{et al}$. 2020, Anatomy of a field-reversed configuration. $\textit{Phys. Plasmas}$, 27, 112508
\bibitem{kako1983equilibrium}Kako M. $\textit{et al}$. 1983, equilibrium of field-reversed configuration with subsidiary coils. $\textit{J. Phys. Soc. Jpn.}$, 52, 3056
\bibitem{kanki1999numerical}Kanki T. $\textit{et al}$. 1999, Numerical simulation of magnetic compression on a field-reversed configuration plasma. $\textit{Phys. Plasmas}$, 6, 4672
\bibitem{gerhardt2006equilibrium}Gerhardt S. P. $\textit{et al}$. 2006, Equilibrium and stability studies of oblate field-reversed configurations in the Magnetic Reconnection Experiment. $\textit{Phys. Plasmas}$, 13, 112508
\bibitem{xie2019gseq}Xie H. S. 2019, MHD Equilibrium Solver GSEQ  (in Chinese). $\textit{ENN}$
\bibitem{johnson1979numerical}Johnson J. L. $\textit{et al}$. 1979, Numerical determination of axisymmetric toroidal magnetohydrodynamic equilibria. $\textit{J. Comput. Phys.}$, 32, 2
\bibitem{kako1983equilibria}Kako M. $\textit{et al}$. 1983, Equilibria of Field-Reversed Configuration with Subsidiary Coils. $\textit{J. Phys. Soc. Jpn.}$, 52, 3056
\bibitem{yambe2009equilibrium}Yambe K. $\textit{et al}$. 2009, Equilibrium of Field-Reversed Configuration Plasma Sustained by Rotating Magnetic Field. $\textit{J. Plasma Fusion Res. SERIES}$, 8, 971
\bibitem{fuentes1995ecmc}Fuentes N. $\textit{et al}$. 1983, ECMC, a portable two-dimensional code for plasma equilibrium computation on coaxial-multiple-coil systems. $\textit{Comput. Phys. Commun.}$, 90, 169
\bibitem{spencer1985experimental}Spencer R. L. $\textit{et al}$. 1985, experimental and computational equilibrium of field-reversed configurations. $\textit{Phys. Fluids}$, 28, 1810
\bibitem{suzuki1999analysis}Suzuki Y. $\textit{et al}$. 1999, Analysis of Averaged B value in Two Dimensional equilibrium of a Field-Reversed configuration with End Mirror Fields. $\textit{J. Plasma Fusion Res. Series.}$, 218
\bibitem{ohkuma2010separatrix}Ohkuma Y. $\textit{et al}$. 2010, Separatrix shape of field-reversed configuration. $\textit{Phys. Plasmas}$, 17, 042502
\bibitem{takahashi2004losses}Takahashi T. $\textit{et al}$. 2004, Losses of neutral beam injected fast ions due to adiabaticity breaking processes in a field-reversed configuration. $\textit{Phys. Plasmas}$, 11, 3131
\end{thebibliography}
\end{document}